\documentclass[aps,amsmath,notitlepage,twocolumn,amssymb,prl,longbibliography]{revtex4-1}

\usepackage{graphicx}
\usepackage{dcolumn}
\usepackage{bm}
\usepackage[colorlinks=true,linkcolor = blue,citecolor=blue,urlcolor=blue,anchorcolor = blue]{hyperref}
\usepackage{wasysym}
\usepackage{stmaryrd}
\usepackage{verbatim}
\usepackage{subfigure}
\usepackage{amsmath}
\usepackage{times}
\usepackage[version=4]{mhchem}
\usepackage{braket}
\usepackage{siunitx}
\usepackage{siunitx}
\usepackage{amssymb}
\newcommand{\RNum}[1]{\uppercase\expandafter{\romannumeral #1\relax}}

\begin{document}

\title{Effective Magnetic Hamiltonian at Finite Temperatures for Rare Earth Chalcogenides}
\author{Zheng\,Zhang$^{1,2}$}
\author{Jianshu\,Li$^{1,2}$}
\author{Weiwei\,Liu$^{1,2}$}
\author{Zhitao\,Zhang$^{3}$}
\author{Jianting\,Ji$^{2}$}
\author{Feng\,Jin$^{2}$}
\author{Rui\,Chen$^{4}$}
\author{Junfeng\,Wang$^{4}$}
\author{Xiaoqun\,Wang$^{5}$}
\author{Jie\,Ma$^{5}$}
\author{Qingming\,Zhang$^{6,2}$}
\email[e-mail:]{qmzhang@ruc.edu.cn}
\affiliation{$^{1}$Department of Physics, Renmin University of China, Beijing 100872, China}
\affiliation{$^{2}$Beijing National Laboratory for Condensed Matter Physics, Institute of Physics, Chinese Academy of Sciences, Beijing 100190, China}
\affiliation{$^{3}$Anhui Province Key Laboratory of Condensed Matter Physics at Extreme Conditions, High Magnetic Field Laboratory, Chinese Academy of Sciences, Hefei 230031 China}
\affiliation{$^{4}$Wuhan National High Magnetic Field Center and School of Physics, Huazhong University of Science and Technology, Wuhan 430074, China}
\affiliation{$^{5}$Department of Physics and Astronomy, Shanghai Jiao Tong University, Shanghai 200240, China}
\affiliation{$^{6}$School of Physical Science and Technology, Lanzhou University, Lanzhou 730000, China}
\date{\today}

\begin{abstract}
	Alkali metal rare-earth chalcogenide \ce{ARECh2} (A=alkali or monovalent metal, RE=rare earth, Ch=O, S, Se, Te), is a large family of quantum spin liquid (QSL) candidates we discovered recently. Unlike \ce{YbMgGaO4}, most members in the family except for the oxide ones, have relatively small crystalline electric-field (CEF) excitation levels, particularly the first ones. This makes the conventional Curie-Weiss analysis at finite temperatures inapplicable and CEF excitations may play an essential role in understanding the low-energy spin physics. Here we considered an effective magnetic Hamiltonian incorporating CEF excitations and spin-spin interactions, to accurately describe thermodynamics in such a system. By taking \ce{NaYbSe2} as an example, we were able to analyze magnetic susceptibility, magnetization under pulsed high fields and heat capacity in a systematic and comprehensive way. The analysis allows us to produce accurate anisotropic exchange coupling energies and unambiguously determine a crossover temperature ($\sim$25 K in the case of \ce{NaYbSe2}), below which CEF effects fade away and pure spin-spin interactions stand out. We further validated the effective picture by successfully explaining the anomalous temperature dependence of electron spin resonance (ESR) spectral width. The effective scenario in principle can be generalized to other rare-earth spin systems with small CEF excitations.
	
\end{abstract}

\maketitle

\emph{Introduction}---Recently rare-earth chalcogenide compounds\cite{liu2018rare} were reported as QSL candidate materials\cite{Anderson1973,ANDERSON1987} and have attracted increasing attention. Extensive studies are focusing on the representative members of \ce{NaYbO2}\cite{bordelon2019field,ding2019gapless}, \ce{NaYbS2}\cite{baenitz2018naybs,sarkar2019quantum}, and \ce{NaYbSe2}\cite{ranjith2019anisotropic,Zhang2020}. Compared to the previous QSL candidate materials like \ce{ZnCu3(OH)6Cl2}\cite{helton2007spin}, \ce{EtMe3Sb[Pd(dmit)2]2}\cite{itou2008quantum}, and \ce{YbMgGaO4}\cite{li2015gapless}, the rare earth family has a perfect triangular spin-lattice. The diversity of the family allows us to control and tune physical properties by various material engineering ways like element replacement or doping. This provides an ideal playground for the study of strongly correlated electron systems with complex magnetic interactions, particularly QSL physics. 
\ce{NaYbSe2} is unique among the family members. Compared to \ce{NaYbO2} and \ce{NaYbS2}, \ce{NaYbSe2} has a Yb-Se-Yb bond angle closer to \ang{90}\cite{jiema_unp}. And high-quality and large single crystal of \ce{NaYbSe2} (5 $\sim$ 15 mm) are already available\cite{gray2003crystal,ranjith2019anisotropic,Zhang2020}. We have pointed out that the smaller energy gap of \ce{NaYbSe2} ($\sim$1.92 eV) suggests a higher possibility of metallization even superconductivity\cite{liu2018rare}. The pressure-induced metallization and superconductivity have been reported very recently\cite{zhang2020pressure,jia2020mott}. 
Particularly, \ce{NaYbSe2} shows a relatively small CEF excitation\cite{Zhang2020,dai2020spinon} compared to other members like \ce{NaYbO2}\cite{ding2019gapless} and \ce{NaYbS2}\cite{baenitz2018naybs}. It means that there exists a delicate competition between CEF excitations and spin-spin interactions at finite temperatures (T$\sim$J) in the system. In this case, one needs to carefully figure out a crossover temperature or energy scale which serves as a boundary wall separating the two parts, and below which we can talk about the low-energy spin Hamiltonian safely and strictly. More generally, the issue more or less exists in most compounds of the family except for the oxide ones. The conventional Curie-Weiss analysis on thermodynamic data has been extensively adopted for rare earth chalcogenides including \ce{NaYbSe2} \cite{liu2018rare,ranjith2019anisotropic,ding2019gapless,baenitz2018naybs,xing2019field,Bastien2020,xing2019synthesis,PhysRevB.99.180401}. The analysis has output valuable basic parameters of the family members. On the other hand, due to the reason discussed above, most of the analysis was established on a non-verified or non-guaranteed basis, if we strictly look into it. We need to investigate the issue accurately and comprehensively.

In this paper, we considered an effective and practical magnetic Hamiltonian which contains both CEF effects and spin-spin interactions. We discussed its physical basis and then applied it to thermodynamic measurements like magnetic susceptibility, magnetization under pulsed high fields, and heat capacity. The thermodynamic data are well simulated and fitted in a mean-field (MF) level, and exchange coupling and other parameters are accurately extracted. 
The CEF contribution to magnetic susceptibility was quantitatively analyzed with the CEF parameters obtained by inelastic neutron scattering (INS) and other measurements in \ce{NaYbSe2}. We further applied the effective scenario to our ESR measurements and successfully explained the anomalous temperature dependence of spectral width. The above simulations self-consistently output a crossover energy scale ($\sim$25 K) below which the system can be strictly described by an anisotropic spin-1/2 picture. This gives us confidence that the feature near 0.8 K can be identified as the strong short-range correlation and quantum fluctuations.

\emph{Samples and Experimental Techniques}---The high-quality single crystals of \ce{NaYbSe2} ($\sim$ 5 mm) were grown by the \ce{NaCl}-flux method\cite{liu2018rare,Zhang2020}. The single crystals were used for magnetic susceptibility, magnetization, pulsed high magnetic field (PHMF), and ESR experiments. The polycrystalline samples of \ce{NaYbSe2}, as well as the nonmagnetic isostructural polycrystalline samples of \ce{NaLuSe2}, were synthesized by the \ce{Se}-flux method\cite{liu2018rare,Zhang2020}. The polycrystalline samples were used for the heat capacity experiments.

$\sim$ 4 mg of \ce{NaYbSe2} single crystals were prepared to make magnetic susceptibility and magnetization measurements. The anisotropic measurements along the c-axis and in the ab-plane were performed using a Quantum Design Physical Property Measurement System (PPMS)  from 1.8 to 300 K under a magnetic field of 0 to14 T.

The high-field magnetization data were obtained from the PHMF, where the magnetic fields were applied parallel to the c-axis and ab-plane of \ce{NaYbSe2} at 4.8 K, 0.8 K and, 0.76 K under 0$\sim$25 T, and $\sim$ 10 mg of \ce{NaYbSe2} single-crystal crystals were used in the measurements.

$\sim$ 1 mg of \ce{NaYbSe2} single crystals were used in the ESR measurements which were performed with a Bruker EMX plus 10/12 continuous-wave spectrometer at X-band frequencies
($f$ $\sim$ 9.39 GHz), where the magnetic fields were applied to the different crystallographic orientations of \ce{NaYbSe2} at 2$\sim$55 K.

The zero-field heat capacity data of polycrystalline samples of \ce{NaYbSe2} ($\sim$ 11 mg) and \ce{NaLuSe2} ($\sim$ 10 mg) were performed using PPMS at 1.8$\sim$100 K and He3 - He4 Dilution Refrigerator(DR) at 50 mK$\sim$4 K.


\emph{Spin-Orbit Coupling, Magnetic Ion Environment and Anisotropic Spin Hamiltonian}---The electron configuration of the \ce{Yb^{3+}} ion is $4f^{13}$\cite{li2015gapless,li2015rare}. For rare-earth ions, $4f$ electrons have strong spin-orbit coupling(SOC)\cite{PhysRevB.94.035107,PhysRevB.96.054445,PhysRevB.97.125105}. The spectral terms of \ce{Yb^{3+}} ion with SOC are $^{2}F_{7/2}$ and $^{2}F_{5/2}$, respectively. Even for free \ce{Yb^{3+}}, the energy level difference between the two states is as high as 1 eV ($\sim $ 10000 K)\cite{zangeneh2019single}. This means that it is enough to focus on the lower $^{2}F_{7/2}$ state at our measurement temperatures. The CEF energy scale in \ce{NaYbSe2} is much smaller than that of SOC and the highest excitation level is about 30 meV($\sim$ 350 K)\cite{Zhang2020}. That is why we took \ce{NaYbSe2} as a representative to investigate the CEF impact on thermodynamics. 


The energy scale of spin-spin exchange coupling in \ce{NaYbSe2} is even lower than 10 K. The CEF excitations will be frozen in the spin ground state at very low temperatures ($T \ll J$). Even in this case, one must also be careful since some external conditions like pulsed high magnetic fields can still activate the CEF effect through the van Vleck mechanism and produce a significant paramagnetic contribution. The solution is to include all the effects in a comprehensive Hamiltonian. A complete Hamiltonian for our case is as follows:
\begin{equation}
\hat{H} = \hat{H}_{0} + \hat{H}_{e-e} + \hat{H}_{SOC} + \hat{H}_{CEF} + \hat{H}_{spin-spin} + \hat{H}_{zeeman}
\end{equation}
where $\hat{H}_{0}$ is the kinetic energy of the electrons and the
electron–nucleus interactions, $\hat{H}_{e-e}$ is the interaction between electrons, $\hat{H}_{SOC}$ is SOC term, $\hat{H}_{CEF}$ is the CEF contribution, $\hat{H}_{spin-spin}$ is the exchange interaction between spins, and $\hat{H}_{zeeman}$ is Zeeman splitting energy produced by the external magnetic field. The first three terms determine the basic shell structure of an isolated $4f$ ion or atom and their energy scales are far away from our measurement conditions (about 60 mK $\sim$ 300 K). The other three terms offer us an effective magnetic Hamiltonian describing thermodynamics at temperatures of interest. Based on the fact that \ce{NaYbSe2} and \ce{YbMgGaO4}\cite{li2015rare} share the same triangular spin-lattice and octahedral configuration, the spin-exchange interaction in \ce{NaYbSe2} can also be described by the anisotropic spin exchange Hamiltonian. The effective magnetic Hamiltonian can be written as\cite{Zhang2020,li2015rare}:
\begin{equation}
\begin{split}
\hat{H}_{eff} &= \hat{H}_{CEF} + \hat{H}_{spin-spin} + \hat{H}_{zeeman} \\
&= \sum_{i}\sum_{m,n} B_{m}^{n} \hat{O}_{m}^{n}  \\
& \quad + \sum_{\left \langle ij \right \rangle} [J_{zz}S_{i}^{z}S_{j}^{z} + J_{\pm}(S_{i}^{+}S_{j}^{-}+S_{i}^{-}S_{j}^{+}) \\
& \quad + J_{\pm \pm}(\gamma_{ij}S_{i}^{+}S_{j}^{+}+\gamma_{ij}^{*}S_{i}^{-}S_{j}^{-}) \\
& \quad -\frac{iJ_{z\pm}}{2}(\gamma_{ij}S_{i}^{+}S_{j}^{z} - \gamma_{ij}^{*}S_{i}^{-}S_{j}^{z} + \left \langle i\longleftrightarrow j \right \rangle)] \\
& \quad -\mu_{0}\mu_{B}\sum_{i} [g_{ab}(h_{x}S_{i}^{x}+h_{y}S_{i}^{y}) + g_{c}h_{c}S_{i}^{z}]
\end{split}
\end{equation}
where $B_{m}^{n}$ is CEF parameters, $J_{zz}$, $J_{\pm}$, $J_{\pm \pm}$, and $J_{z\pm}$ are anisotropic spin-exchange parameters, the phase factor $\gamma_{ij} = 1, e^{i2\pi/3}, e^{-i2\pi/3}$ for the nearest neighbor (NN) interactions, and $g_{ab}$ and $g_{c}$ ($\sim$ 3.1 and 0.97)\cite{Zhang2020} represent the Lande factors in the directions of the ab-plane and the c-axis, respectively. The effective Hamiltonian allows us to quantitatively estimate the contributions from CEF, spin-exchange interaction, and the external magnetic field, and accurately fit the fundamental parameters in the Hamiltonian.

\begin{figure}[t]
	\includegraphics[scale=0.92]{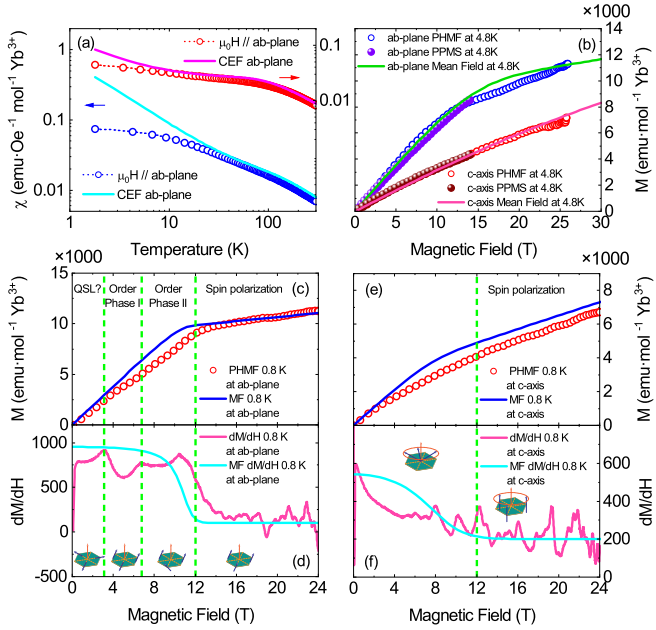}
	\caption{\label{fig:epsart} Magnetic measurements under pulsed high magnetic fields (PHMF). Open and closed circles are experimental data and solid lines except for the cyan ones in (d) and (f) are mean-field (MF) simulations. (a) Temperature-dependent magnetic susceptibility. (b) Magnetization under PHMF at 4.8 K. The data are calibrated with PPMS up to 14 T. (c) and (e) Magnetization under PHMF in the ab-plane and along the c-axis at 0.8 K, respectively. Their corresponding derivatives are shown in (d) and (f). The insets in (d) and (f) show the possible spin configurations under different magnetic fields.}
\end{figure}

\begin{figure*}[t]
	\includegraphics[scale=1]{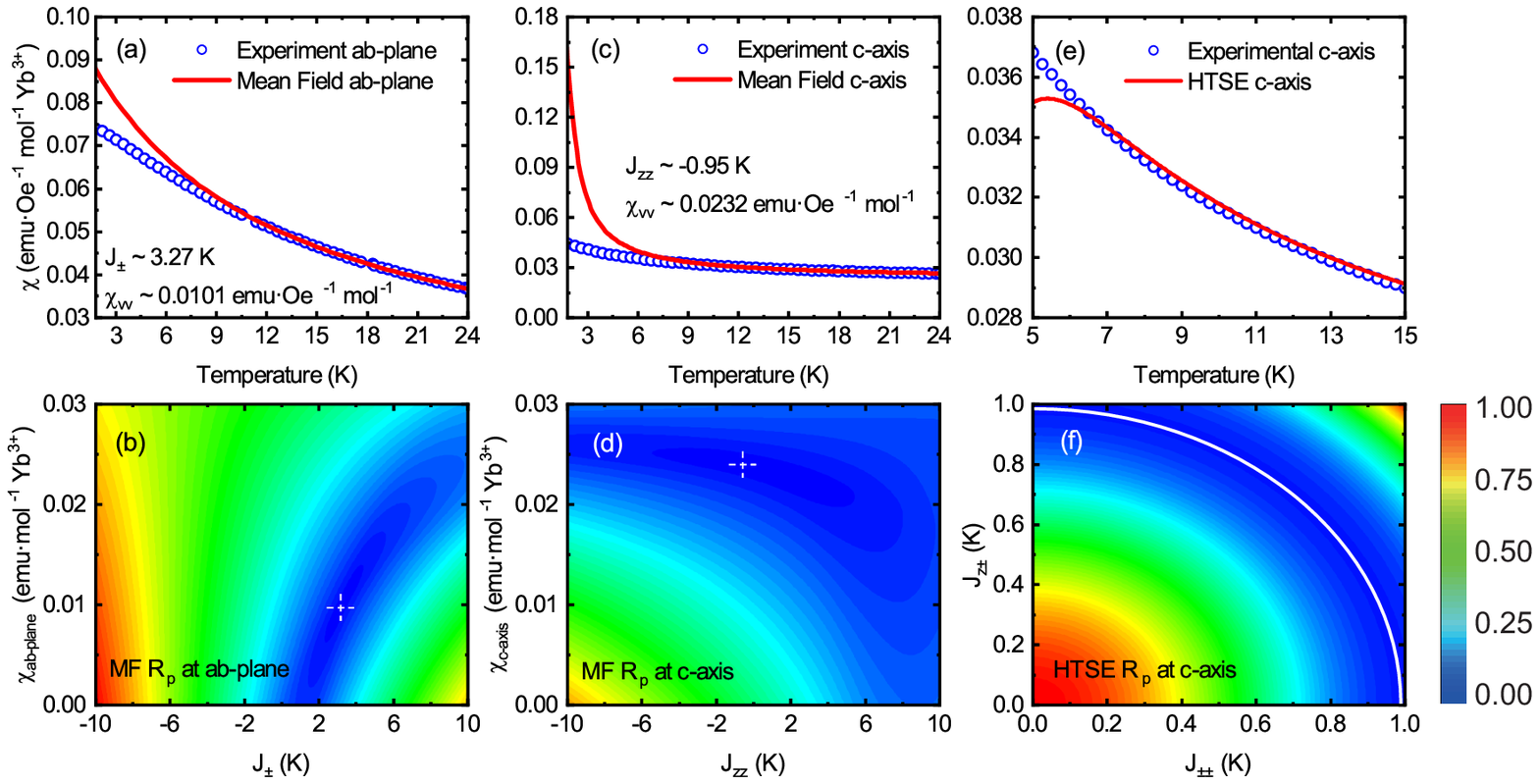}
	\caption{\label{fig:epsart} Magnetic susceptibility and anisotropic exchange parameters. (a), (c) and (e) Open circles and solid lines are experimental data and mean-field (MF) or high-temperature series expansion (HTSE) simulations, respectively. (b) and (d) The deviation $R_{p}$ of the experimental susceptibility from the MF calculation results in the ab-plane and the c-axis. The white crosses indicate optimal values. (f) The deviation $R_{p}$ of the experimental susceptibility from the HTSE calculation results in the c-axis. The white dashed line marks the possible optimal values. (b), (d) and (f) share the same color scale in the lower right corner.}
\end{figure*}

\emph{Susceptibility, Magnetization and Pulsed High Magnetic Fields}---The experimentally determined CEF excitation levels in \ce{NaYbSe2} are located at 15.79 meV, 24.33 meV and, 30.53 meV, respectively\cite{Zhang2020,dai2020spinon}. It is expected that the CEF excitations will significantly contribute to thermodynamics at finite temperatures. We calculated the CEF contributions to magnetic susceptibility in the ab-plane and along the c-axis (shown in Fig. 1(a))\cite{SI}. A characteristic temperature of $\sim$25 K emerges from the comparison between experimental and calculation results. The calculated curves well match the experimental data above the temperature, while the calculations significantly deviate from the experimental ones below the temperature. The temperature is an energy scale below which the CEF contributions are negligible and spin-spin interactions become dominant. It should be noted that the general orbital diamagnetism in any material is responsible for the slight difference between the calculated and measured results above 25 K.

If the effective scenario works, one can anticipate that the CEF effect frozen below the characteristic temperature can be reactivated by high magnetic fields through the van Vleck mechanism. That is what one can see in Fig. 1(b), in which magnetization under PHMF at 4.8 K is measured and the self-consistent MF simulations including both CEF excitations and spin-spin interactions, are in agreement with the experiments. More specifically, the agreement along the c-axis seems good while it shows a deviation in ab-plane under higher magnetic fields ($\textgreater$ 14 T). There are several possibilities for the deviation. One reason is the increasing instrumental vibration under higher fields. Another reason may be that the measurement temperature of 4.8 K is already close to the scale of spin exchange coupling, and the strong fluctuations may reduce the reliability of MF approximation. It should be noted that the PHMF data have been calibrated using PPMS measurements.

The magnetization under PHMF at a lower temperature of 0.8 K is presented in Fig. 1(c) and 1(e), and the MF self-consistent calculations have also been carried out. The calculated results follow the experimental data. When further checking the derivatives of magnetization (Fig. 1(d) and Fig. 1(f)), one may catch some subtle signs of spin transitions between the spin states induced by magnetic fields, which have been revealed by neutron scattering\cite{ma2020spin,jiema_unp,bordelon2019field,dai2020spinon}. With increasing temperatures, the spin system sequentially undergoes the phases of the spin liquid, up-up-down, spin-flop, and spin-flip, which corresponds to three spin transitions at 3.1 T, 6.8 T and 12 T, respectively. The system seems to be fully polarized above $\sim$12 T.

The effective magnetic Hamiltonian allows extracting spin exchange and other fundamental parameters in a more accurate way. We were able to do the fitting by using a self-consistent equation based on MF approximation\cite{SI} (Fig. 2). The temperature range for fitting is safely selected to be 15$\sim$24 K, according to the above analysis. The fitting gives the spin exchange parameter $J_{\pm}$ of 3.27 K and $J_{zz}$ of -0.95 K. And we have also obtained the paramagnetic susceptibility which related to the van Vleck mechanism contributed by the CEF ground state, and the values in the ab-plane and along the c-axis are 0.0101 ${\rm emu\cdot Oe^{-1} \cdot mol^{-1}}$ and 0.0232 ${\rm emu\cdot Oe^{-1} \cdot mol^{-1}}$, respectively. It should be noted that the self-consistent calculations begin to deviate from the experimental data below 10 K (Fig. 2(a) and 2(c)). The deviation can be explained by the fact that the contributions from $J_{\pm\pm}$ and $J_{z\pm}$ terms are smeared out in the MF approximation.

The other two anisotropic spin exchange interactions $J_{\pm\pm}$ and $J_{z\pm}$ are extracted from high-temperature series expansion (HTSE), which gives the following expression of magnetic susceptibility along c-axis\cite{li2015rare}:
\begin{small}
	\begin{equation}
	\chi _{c} = \frac{{{\mu _0}g_{c}^2\mu _B^2}}{{4{k_B}T}}\left( {1 - \frac{{3{J_{zz}}}}{{2{k_B}T}} - \frac{{3J_ \pm ^2 + J_{ \pm  \pm }^2 + J_{z \pm }^2}}{{2k_B^2{T^2}}} + \frac{{15J_{zz}^2}}{{8k_B^2{T^2}}}} \right)
	\end{equation}
\end{small}
We made the HTSE fitting for magnetic susceptibility along the c-axis from 5 to 15 K (Fig. 2(e)). The HTSE simulations effectively suppress the divergent trend in the MF calculations below 10 K, as anisotropic spin exchange interactions $J_{\pm\pm}$ and $J_{z\pm}$ are taken into account in the HTSE method. Fig. 2(f) further presents the parameters space for HTSE and the while line marks the optimal values for  $J_{\pm\pm}$ and $J_{z\pm}$, i.e. $|J_{\pm\pm}|^{2}+|J_{z\pm}|^{2}=0.965$ $K^{2}$. The square form stems from the expression (3) itself. In Fig. 2, we plot the normalized deviation of the experimental results from the MF or HTSE calculation one
\begin{equation}
	R_{p} = Norm\left[\frac{|\chi_{exp}-\chi_{the}|}{\chi_{exp}}\right]
\end{equation}

\begin{figure}[t]
	\includegraphics[scale=0.34]{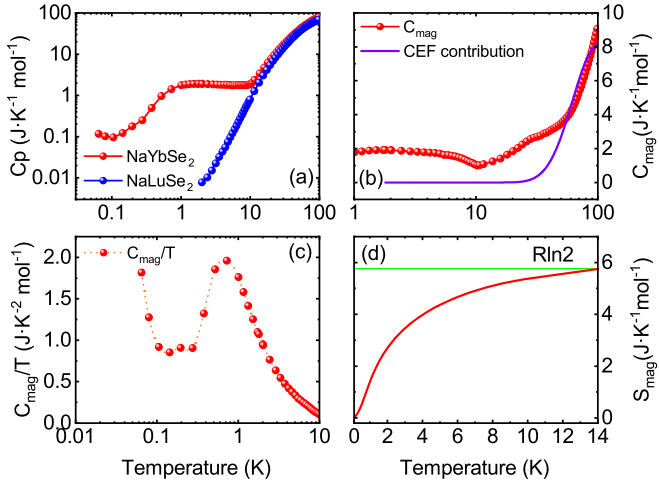}
	\caption{\label{fig:epsart}(a)Heat capacity of \ce{NaYbSe2} and \ce{NaLuSe2}. (b) Magnetic heat capacity of \ce{NaYbSe2}. The red closed circles are experimental data and the purple solid line is given by CEF calculations. (c) Temperature dependence of magnetic heat capacity of \ce{NaYbSe2} at ultra-low temperatures. (d) Magnetic entropy of \ce{NaYbSe2}. }
\end{figure}

\emph{Zero-Field Heat Capacity}---In the effective picture, it can be expected that the weaker CEF in \ce{NaYbSe2} will also contribute to magnetic heat capacity. That motivates us to examine heat capacity of \ce{NaYbSe2} under zero field. We measured the heat capacity of the non-magnetic control sample \ce{NaLuSe2} to experimentally determine the phonon contribution to the heat capacity. As shown in Fig. 3(a), the heat capacity of \ce{NaLuSe2} follows the Debye model well\cite{SI} and is less than 0.01 $\rm J\cdot K^{-1} mol^{-1}$ below 2 K. We obtained magnetic heat capacity by extracting the phonon part given by \ce{NaLuSe2}. It should be emphasized here that the extracted magnetic heat capacity is relatively accurate since the contribution from phonons is experimentally rather than theoretically determined\cite{li2015gapless}. 

Then we calculated the CEF contribution to magnetic heat capacity in \ce{NaYbSe2}\cite{SI} (Fig. 3(b)). There is a reasonable agreement between experiments and simulations. We can further identify that the magnetic heat capacity at higher temperatures (\textgreater 50 K) is dominated by CEF excitations. More importantly, the CEF contribution to heat capacity goes down to zero below 25 K, consistent with what we have seen in magnetic susceptibility in Fig. 1. Again it suggests a characteristic energy scale of 25 K, below which the CEF influence fades away and the system can be safely described by an effective spin-1/2 picture. 

The magnetic heat capacity at ultra-low temperatures is shown in Fig .3 (c)\cite{liu2018rare}. A broad peak appears in the temperature range of 0.5-2 K. The similar broad peaks in magnetic heat capacity have been observed in \ce{NaYbO2}\cite{bordelon2019field}, \ce{NaYbS2}\cite{baenitz2018naybs}, and many other quantum spin liquid candidates\cite{li2015gapless}. Considering that the family is almost disorder or impurity free and the CEF contribution has been ruled out, it is reasonable to attribute the peak to strong spin fluctuations. And the anisotropic spin-exchange parameters $J_{\pm\pm}$ and $J_{z\pm}$ ($\sim$1 K) coincide with the peak position. The rapid increase below 0.1 K is related to the contribution from nuclei. 

By integrating magnetic heat capacity, we obtained magnetic entropy (Fig. 3(d)) which verifies the validity of the effective spin-1/2 picture below 25 K. Compared to \ce{YbMgGaO4} ($\sim$50 K)\cite{li2015rare,SI}, \ce{NaYbSe2} has a lower characteristic temperature below which the system enters into an effective spin-1/2 state, due to the smaller first CEF excitation.

\begin{figure*}[t]
	\includegraphics[scale=0.22]{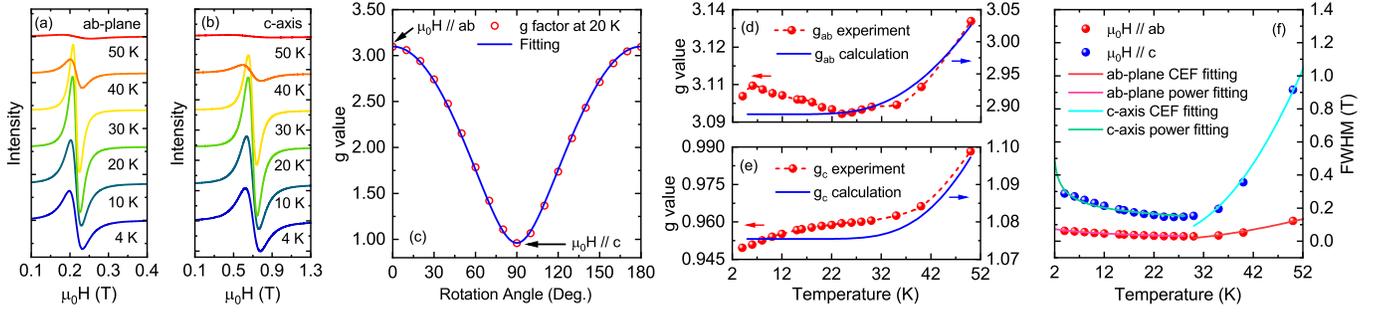}
	\caption{\label{fig:epsart} (a) and (b) Differential ESR spectra of \ce{NaYbSe2} in the ab-plane and along the c-axis. (c) Rotation-angle dependence of g factors at 20 K. (d) and (e) Temperature-dependent of g factors in the ab-plane and along the c-axis. The red closed circles are experimental data and the blue solid lines are CEF calculations. (f) ESR full width at half maximum (FWHM) of \ce{NaYbSe2} vs. the temperature. The  closed circles are experimental data and the solid lines are obtained by different fitting schemes.}
\end{figure*}

\emph{Electron Spin Resonance}---The CEF effect not only affects thermodynamics but also has an impact on spectral excitations. Here, we studied ESR spectra of \ce{NaYbSe2}, which are essentially different from those of \ce{YbMgGaO4}\cite{li2015rare} and \ce{ZnCu3(OH)6Cl2}\cite{zorko2008dzyaloshinsky}. The differential ESR spectra in \ce{NaYbSe2} are significantly broadened above 50 K. This is similar to the case of \ce{NaYbS2}\cite{sichelschmidt2019electron}. The rotation-angle dependence of the obtained g factors at 20 K shows a strong anisotropy. The angle dependence can be described by the following formula:
\begin{equation}
g(\theta) = \sqrt{g_{ab-plane}^{2}cos^{2}(\theta) + g_{c-axis}^{2}sin^{2}(\theta)}
\end{equation}
where $\theta$ is the angle between the magnetic field and the c-axis. Then we obtained $g_{ab} = 3.10$  and $g_{c} = 0.96$. The values are consistent with those extracted from inelastic neutron scattering experiments\cite{Zhang2020}. Generally, the g factors given by ESR experiments are more direct and accurate. The temperature dependence of $g_{ab-plane}$ and $g_{c-axis}$ is shown in Fig. 4(d) and 4(e) and is compared to the CEF calculations. Again we meet the characteristic temperature of $\sim$25 K, below which the g-factors both in the ab-plane and along the c-axis exhibit little temperature dependence. But the g factors rapidly increase with the temperature above 25 K. We can develop a quantitative understanding using the following formula involving CEF excitations:
\begin{gather}
\ket{\psi_{+}} = \frac{1}{Z}\sum_{i=0,1,2,3} exp(-E_{i}/k_{B}T)\ket{\psi_{i,+}}\\
\ket{\psi_{-}} = \frac{1}{Z}\sum_{i=0,1,2,3} exp(-E_{i}/k_{B}T)\ket{\psi_{i,-}}\\
g_{ab-plane} = g_{j}|\bra{\psi_{\pm}}m_{J_{\pm}}\ket{\psi_{\mp}}| \\
g_{c-axis} = 2g_{j}|\bra{\psi_{\pm}}m_{J_{z}}\ket{\psi_{\pm}}|
\end{gather}
where $Z$ is the partition function, $g_{j}=8/7$,  $E_{i}$ and $\ket{\psi_{\pm}}$ are Lande factor, the CEF energy level, and the CEF eigenstate\cite{Zhang2020,SI}. The simulation curves coincide with the experimental one (Fig. 4(d) and 4(e))). The difference between experiments and simulations may be explained by the fact that the CEF eigenstates adopted here come from INS experiments while the g factors here are given by ESR measurements. It is worth noting that the g factors slightly increase with temperature, especially above 25 K. As the temperature increases, more and more magnetic ions are excited to higher CEF states. And this eventually causes a subtle change in the g factor.

The temperature dependence of FWHM of ESR spectra in \ce{NaYbSe2}, is shown in Fig. 4(f). The FWHM data in the ab-plane and along the c-axis share a similar temperature evolution, but with distinct amplitudes. This is simply an indication of strong anisotropy. We can discuss FWHM in the two temperature regimes separated by the characteristic temperature. For $T > $ 25 K, the rapid increase of FWHM with temperature is related to the CEF excitations. Using the following formula, we can estimate the energy gap ($\Delta E$) between the CEF ground state and the first excited state in \ce{NaYbSe2}\cite{sichelschmidt2019electron,Sichelschmidt2020}.
\begin{equation}
\mu_{0}\Delta H \simeq \frac{1}{exp(\Delta E/T)-1}
\end{equation}
This gives the CEF excitation gap of 133 K (ab-plane) and 169 K (c-axis). The result is well consistent with that measured by INS experiments ($\sim$ 180 K)\cite{Zhang2020}.
Below 25 K, FWHM slowly increases with decreasing temperatures. This is due to anisotropic spin-exchange interactions. $\mu_{0}\Delta H$ can be fitted to the following empirical formula\cite{seehra1970critical,reynaud2001orbital,sichelschmidt2019electron}:
\begin{equation}
\mu_{0}\Delta H \simeq \frac{A}{(T-T_{\theta})^{p}}
\end{equation}
where $A$ is the proportionality coefficient, $p$ is the power coefficient, and $T_{\theta}$ is the Weiss temperature, and $T_{\theta}^{ab}=-3J_{\pm}$ and $T_{\theta}^{c}=-3J_{zz}/2$\cite{li2015rare}. $J_{\pm}$ and $J_{zz}$ have been obtained by making MF fitting for magnetic susceptibility discussed above. The fitting gives $A=0.4626$, $p=0.7458$ in the ab-plane and $A=0.4126$, $p=0.3024$ along the c-axis. Since the FWHM is proportional to the local magnetic susceptibility, the magnetic susceptibility in the ab-plane and along the c-axis is very similar to ESR FWHM in both the trend and the intensity. This is reflected in the similarity between the empirical formula (11) and Curie-Weiss law, regardless of the power coefficient $p$. The power coefficient $p$ of less than 1 stems from spin-spin correlation below 25 K and strong magnetic anisotropy. Both factors push the magnetic behavior at low temperatures away from the picture of simply isolated ions (Curie-Weiss law). The strong anisotropy is again demonstrated by the distinct power coefficients $p$ in the ab-plane and along the c-axis.

The broadening of ESR linewidth at low temperatures has also been observed in other spin-frustrated materials, such as \ce{ZnCu3(OH)6Cl2}\cite{zorko2008dzyaloshinsky} and \ce{LiNiO2}\cite{reynaud2001orbital}. It should be pointed out that the temperature evolution of FWHM in \ce{NaYbSe2} is substantially different from that of \ce{YbMgGaO4}\cite{li2015rare}. No obvious broadening of ESR FWHM in \ce{YbMgGaO4} can be seen when varying temperatures. This is explained by the higher CEF levels of \ce{YbMgGaO4} ($\textgreater$300 K)\cite{li2017crystalline}. In addition, the spin-exchange parameters of \ce{YbMgGaO4} is smaller (less than 1 K)\cite{li2015rare}. Therefore, spin-exchange interactions have little effect on ESR linewidth at relatively high temperatures.

\emph{Summary}---We conducted a comprehensive analysis of thermodynamic data and the ESR spectra of \ce{NaYbSe2}. We considered an effective magnetic Hamiltonian involving strong SOC and CEF effects. At the MF level, we can well understand  the thermodynamic data of \ce{NaYbSe2}. The analysis determines a characteristic temperature of $\sim$25 K. The CEF effect makes a major contribution to thermodynamics above the temperature. On the other hand, the spin-spin correlation begins to play a major role below the temperature. The consideration allows us to separate the CEF contribution from thermodynamic data and accurately determine the fundamental parameters of the spin system. Furthermore, we applied the effective scenario to ESR measurements and successfully explained the unusual temperature evolution of FWHM. The present study provides a realistic scheme to include the CEF influence in a comprehensive way and may be generalized to other rare-earth systems.

\emph{Acknowledgments}---This work was supported by the National Key Research and Development Program of China (2017YFA0302904 \& 2016YFA0300500) and the NSF of China (11774419 \& U1932215). A portion of this work was performed on the Steady High Magnetic Field Facilities, High Magnetic Field Laboratory, CAS. The CEF calculations of magnetic susceptibility is based on our python program package MagLab\cite{SI}, which is designed for the fitting, calculations, and analysis of the CEF magnetism, and will be continuously updated. The MF analysis and fitting of magnetic susceptibility, magnetization and ESR data at low temperatures are based on MathWorks$^\circledR$ MATLAB$^\circledR$ software (Academic License for Renmin University of China).

%

\end{document}


\title{Supplementary Information: 
Effective Magnetic Hamiltonian at Finite Temperatures for Rare Earth Chalcogenides}
\author{Zheng\,Zhang$^{1,2}$}
\author{Jianshu\,Li$^{1,2}$}
\author{Weiwei\,Liu$^{1,2}$}
\author{Zhitao\,Zhang$^{3}$}
\author{Jianting\,Ji$^{2}$}
\author{Feng\,Jin$^{2}$}
\author{Rui\,Chen$^{4}$}
\author{Junfeng\,Wang$^{4}$}
\author{Xiaoqun\,Wang$^{5}$}
\author{Jie\,Ma$^{5}$}
\author{Qingming\,Zhang$^{6,2}$}
\email[e-mail:]{qmzhang@ruc.edu.cn}
\affiliation{$^{1}$Department of Physics, Renmin University of China, Beijing 100872, China}
\affiliation{$^{2}$Beijing National Laboratory for Condensed Matter Physics, Institute of Physics, Chinese Academy of Sciences, Beijing 100190, China}
\affiliation{$^{3}$Anhui Province Key Laboratory of Condensed Matter Physics at Extreme Conditions, High Magnetic Field Laboratory, Chinese Academy of Sciences, Hefei 230031 China}
\affiliation{$^{4}$Wuhan National High Magnetic Field Center and School of Physics, Huazhong University of Science and Technology, Wuhan 430074, China}
\affiliation{$^{5}$Department of Physics and Astronomy, Shanghai Jiao Tong University, Shanghai 200240, China}
\affiliation{$^{6}$School of Physical Science and Technology, Lanzhou University, Lanzhou 730000, China}
\date{\today}

\begin{abstract}
We present here:

1. Atomic orbital wave function of \ce{Yb^{3+}}

2. Calculation method of crystalline electric-field (CEF) susceptibility

3. The mean field approximation of anisotropic spin Hamiltonian.

4. The heat capacity curve of \ce{NaLuSe2} and the Debye fitting at low temperature.

5. MagLab, a CEF analysis software package developed based on python
\end{abstract}

\maketitle

\section{Wave Functions}
\subsection{Point group symmetry}
In \ce{NaYbSe2}, the magnetic \ce{Yb^{3+}} cation and the surrounding ligand \ce{Se^{2-}} form a local structure with $D_{3d}$ point group symmetry. The 7 $4f$ atomic orbitals of \ce{Yb^{3+}} are expressed as follows:
\begin{align}
&f_{z(5z^{2}-3r^{2})}=f_{z^{3}}=\frac{1}{4}\sqrt{\frac{7}{\pi}}\left(5cos^{2}\theta-3cos\theta\right)=\frac{1}{4}\sqrt{\frac{7}{\pi}}\frac{z}{r^{3}}\left(5z^{3}-3r^{2}\right)\\
&f_{x(5z^{2}-r^{2})}=f_{xz^{2}}=\frac{1}{8}\sqrt{\frac{42}{\pi}}sin\theta\left(5cos^{2}\theta-1\right)cos\psi=\frac{1}{8}\sqrt{\frac{42}{\pi}}\frac{x}{r^{3}}\left(5z^{2}-r^{2}\right)\\
&f_{y(5z^{2}-r^{2})}=f_{yz^{2}}=\frac{1}{8}\sqrt{\frac{42}{\pi}}sin\theta(5cos^{2}\theta-1)sin\psi=\frac{1}{8}\sqrt{\frac{42}{\pi}}\frac{y}{r^{3}}\left(5z^{2}-r^{2}\right)\\
&f_{z(x^{2}-y^{2})}=\frac{1}{4}\sqrt{\frac{105}{\pi}}sin^{2}\theta cos\theta cos2\psi = \frac{1}{4}\sqrt{\frac{105}{\pi}}\frac{z}{r^{3}}\left(x^{2}-y^{2}\right)\\
&f_{xyz}=\frac{1}{4}\sqrt{\frac{105}{\pi}}sin^{2}\theta cos\theta sin2\psi = \frac{1}{2}\sqrt{\frac{105}{pi}}\frac{xyz}{r^{3}}\\
&f_{x(x^{2}-3y^{2})}=\frac{1}{8}\sqrt{\frac{70}{\pi}}sin^{3}\theta cos3\psi = \frac{1}{8}\sqrt{\frac{70}{\pi}}\frac{x}{r^{3}}\left(x^2-3y^{2}\right)\\
&f_{y(3x^{2}-y^{2})}=\frac{1}{8}\sqrt{\frac{70}{\pi}}sin^{3}\theta sin3\psi = \frac{1}{8}\sqrt{\frac{70}{\pi}}\frac{y}{r^{3}}\left(3x^{2}-y^{2}\right)
\end{align}
The irreducible representation of the $D_{3d}$ point group is shown in TABLE I\cite{dresselhaus2007group}.
\begin{table}
	\caption{\label{tab:table1}Character table for point group $D_{3d}$}
	\begin{ruledtabular}
		\begin{tabular}{cccccccc}
			$D_{3d}$ & E & 2$C_{3}$ & 3$C_{2}^{'}$ & i & 2$S_{6}$ & 3$\sigma_{d}$ & cubic functuions \\
			\hline
			$A_{1g}$ & +1 & +1 & +1 & +1 & +1 & +1 & - \\
			\hline
			$A_{2g}$ & +1 & +1 & -1 & +1 & +1 & -1 & - \\
			\hline
			$E_{g}$ & +2 & -1 & 0 & +2 & -1 & 0 & - \\
			\hline
			$A_{1u}$ & +1 & +1 & +1 & -1 & -1 & -1 & $x(x^{2}-3y^{2})$ \\
			\hline
			$A_{2u}$ & +1 & +1 & -1 & -1 & -1 & +1 & $y(3x^{2}-y^{2})$, $z^{3}$, $z(x^{2}+y^{2})$ \\
			\hline
			$E_{u}$ & +2 & -1 & 0 & -2 & +1 & 0 & $(xz^{2}, yz^{2})$, $[xyz, z(x^{2}-y^{2})]$, $[x(x^{2}+y^{2}), y(x^{2}+y^{2})]$
		\end{tabular}
	\end{ruledtabular}
\end{table}
The irreducible representation of 7 $4f$ orbits is shown in TABLE II.
\begin{table}
	\centering
	\small
	\caption{\label{tab:table1}Irreducible representation of $4f$ orbits under $D_{3d}$ point group}
		\begin{tabular}{cc}
			\hline
			Irreducible representation & wave functions \\
			\hline
			$A_{1u}$ & $f_{x(x^{2}-3y^{2})}$ \\
			\hline
			$A_{2u}$ & $f_{z^{3}}$, $f_{y(3x^{2}-y^{2})}$\\
			\hline
			$E_{u}$ & $(f_{xz^{2}}, f_{yz^{2}})$, $(f_{xyz}, f_{z(x^{2}-y^{2})})$ \\
			\hline 
		\end{tabular}
\end{table}
\subsection{Spin-Orbit Coupling (SOC), Magnetic Ion Environment}
Considering that \ce{Yb^{3+}} has strong spin-orbit coupling(SOC) in \ce{NaYbSe2}, cations with $^{2}F$ spectral term split into $^{2}F_{7/2}$ and $^{2}F_{5/2}$ spectral terms. In the crystalline electric-field(CEF) environment, the wave functions of the two spectral terms are shown in   expression (8)\cite{Ranjith2019} and expression (9)
\begin{equation}
^2{F_{7/2}}\left\{ \begin{array}{l}
\left| {\varphi _1^ \pm } \right\rangle  =  \mp {\alpha _1}\left| {\frac{7}{2}, \pm \frac{7}{2}} \right\rangle  + {\alpha _2}\left| {\frac{7}{2}, \pm \frac{1}{2}} \right\rangle  \pm {\alpha _3}\left| {\frac{7}{2}, \pm \frac{5}{2}} \right\rangle \\
\\
\left| {\varphi _2^ \pm } \right\rangle  = {\beta _1}\left| {\frac{7}{2}, \pm \frac{7}{2}} \right\rangle  \pm {\beta _2}\left| {\frac{7}{2}, \pm \frac{1}{2}} \right\rangle  + {\beta _3}\left| {\frac{7}{2}, \mp \frac{5}{2}} \right\rangle \\
\\
\left| {\varphi _3^ \pm } \right\rangle  = \left| {\frac{7}{2}, \pm \frac{3}{2}} \right\rangle \\
\\
\left| {\varphi _4^ \pm } \right\rangle  =  \mp {\gamma _1}\left| {\frac{7}{2}, \pm \frac{7}{2}} \right\rangle  - {\gamma _2}\left| {\frac{7}{2}, \pm \frac{1}{2}} \right\rangle  \pm {\gamma _3}\left| {\frac{7}{2}, \mp \frac{5}{2}} \right\rangle 
\end{array} \right.
\end{equation}
\begin{equation}
	^2{F_{5/2}}\left\{ \begin{array}{l}
	\left| {\varphi _5^ \pm } \right\rangle  =  \mp \left| {\frac{5}{2}, \mp \frac{3}{2}} \right\rangle \\
	\\
	\left| {\varphi _6^ \pm } \right\rangle  =  - {\chi _1}\left| {\frac{5}{2}, \pm \frac{1}{2}} \right\rangle  \mp {\chi _2}\left| {\frac{5}{2}, \mp \frac{5}{2}} \right\rangle \\
	\\
	\left| {\varphi _7^ \pm } \right\rangle  = {\zeta _1}\left| {\frac{5}{2}, \pm \frac{5}{2}} \right\rangle  \pm {\zeta _2}\left| {\frac{5}{2}, \mp \frac{1}{2}} \right\rangle 
	\end{array} \right.
\end{equation}
At the same time, the coefficient of the equation satisfies the normalization condition, that is, $|\alpha_{1}|^{2} + |\alpha_{2}|^{2} + |\alpha_{3}|^{2} = 1$, $|\beta_{1}|^{2}+|\beta_{2}|^{2}+|\beta_{3}|^{2}=1$, $|\gamma_{1}|^{2}+|\gamma_{2}|^{2}+|\gamma_{3}|^{2}=1$, $|\chi_{1}|^{2} + |\chi_{2}|^{2} = 1$ and $|\xi_{1}|^{2} + |\xi_{2}|^{2} = 1$. These coefficients can be obtained by the experimental method of inelastic neutron scattering(INS) or by the method of quantum chemical calculation. The wave function expression of $^{2}F_{7/2}$ spectral terms obtained from INS experiment is as follows\cite{Zhang2020a}:
\begin{equation}
	^2{F_{7/2}}\left\{ \begin{array}{l}
	\left| {\varphi _1^ \pm } \right\rangle  = 0.8019\left| {\frac{7}{2}, \pm \frac{5}{2}} \right\rangle  \pm 0.1368\left| {\frac{7}{2}, \mp \frac{1}{2}} \right\rangle  - 0.5693\left| {\frac{7}{2}, \mp \frac{7}{2}} \right\rangle \\
	\\
	\left| {\varphi _2^ \pm } \right\rangle  =  - 0.1396\left| {\frac{7}{2}, \pm \frac{3}{2}} \right\rangle  \pm 0.9902\left| {\frac{7}{2}, \mp \frac{3}{2}} \right\rangle \\
	\\
	\left| {\varphi _3^ \pm } \right\rangle  =  - 0.4716\left| {\frac{7}{2}, \pm \frac{5}{2}} \right\rangle  \pm 0.7307\left| {\frac{7}{2}, \mp \frac{1}{2}} \right\rangle  - 0.4887\left| {\frac{7}{2}, \mp \frac{7}{2}} \right\rangle \\
	\\
	\left| {\varphi _4^ \pm } \right\rangle  =  - 0.3524\left| {\frac{7}{2}, \pm \frac{5}{2}} \right\rangle  \mp 0.6665\left| {\frac{7}{2}, \mp \frac{1}{2}} \right\rangle  - 0.6565\left| {\frac{7}{2}, \mp \frac{7}{2}} \right\rangle 
	\end{array} \right.
\end{equation}
Therefore, the wave function of magnetic ion \ce{Yb^{3+}} in \ce{NaYbSe2} with and without spin-orbit coupling is shown in Fig. S1(b) and (c).
\begin{figure}[t]
	\includegraphics[scale=0.4]{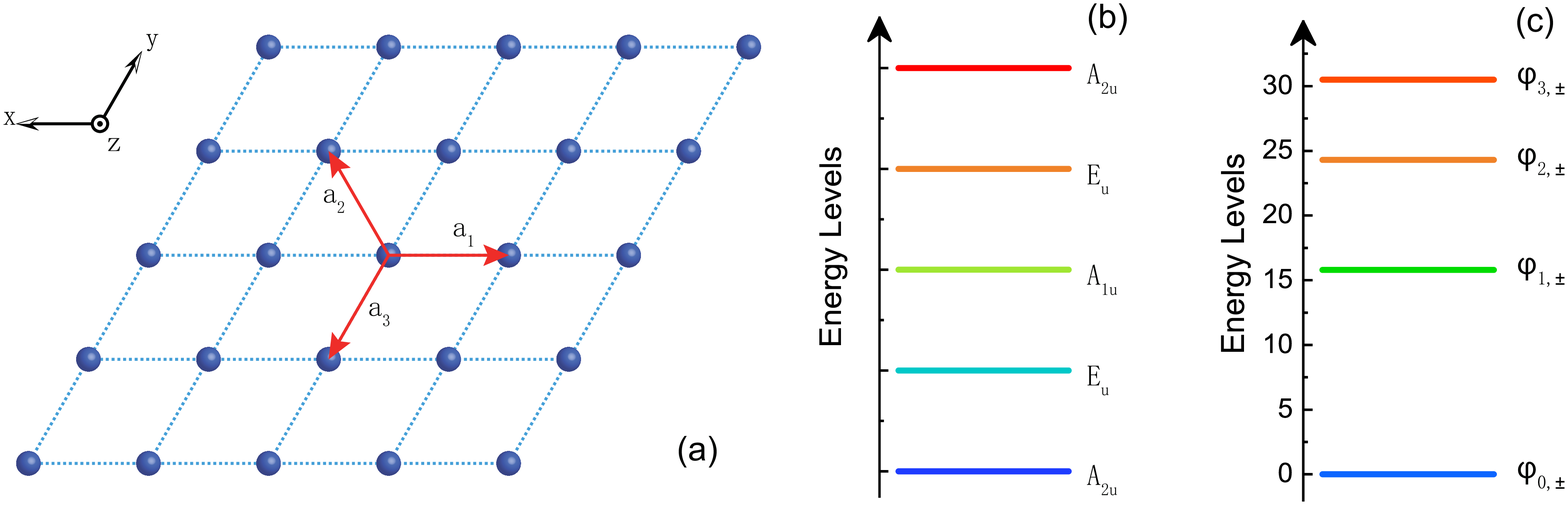}
	\caption{\label{fig:epsart}(a) Triangular plane structure composed of \ce{Yb^{3+}} in \ce{NaYbSe2}.
		(b) Schematic diagram of the energy levels of $4f$ orbits without spin-orbit coupling under the symmetry of $D_{3d}$ point group. (c) CEF
		energy levels of \ce{Yb^{3+}} cation with strong spin-orbit coupling.}
\end{figure}

\section{The contribution of CEF to susceptibility and specific heat capacity}
The Hamiltonian of CEF can be written as follows:
\begin{equation}
	H_{CEF} = \sum_{i} B_{2}^{0}O_{2}^{0} + B_{4}^{0}O_{4}^{0} + B_{4}^{3}O_{4}^{3} + 
	B_{6}^{0}O_{6}^{0} + B_{6}^{3}O_{6}^{3} + B_{6}^{6}O_{6}^{6} -\mu_{0}\mu_{B} \sum_{i}h_{x}J_{x} + h_{y}J_{y} + h_{z}J_{z}
\end{equation}
The corresponding eigenenergy $E_{i}$ and eigenstate $\left| {{\varphi _i}} \right\rangle$ can be obtained by diagonalizing the Hamiltonian. At this time, the susceptibility of the CEF in the c-axis direction and the ab-plane is expressed as follows\cite{Arnold2014,Li2020}:
\begin{equation}
	\chi _{c - axis}^{CEF} = \frac{{{\mu _B}{g_c}{N_A}\sum\limits_{i = 1}^8 {\exp \left( { - \frac{{{E_i}}}{{{k_B}T}}} \right)\left\langle {{\varphi _i}} \right|{J_z}\left| {{\varphi _i}} \right\rangle } }}{{{h_z}\sum\limits_{i = 1}^8 {\exp \left( { - \frac{{{E_i}}}{{{k_B}T}}} \right)} }}
\end{equation}
\begin{equation}
	\chi _{ab - plane}^{CEF} = \frac{{{\mu _B}{g_{ab}}{N_A}\sum\limits_{i = 1}^8 {\exp \left( { - \frac{{{E_i}}}{{{k_B}T}}} \right)\left\langle {{\varphi _i}} \right|{J_x}\left| {{\varphi _i}} \right\rangle } }}{{{h_z}\sum\limits_{i = 1}^8 {\exp \left( { - \frac{{{E_i}}}{{{k_B}T}}} \right)} }}
\end{equation}
where $g_{c}$ and $g_{ab}$ are g factor along the c-axis and in the ab-plane, respectively.

The calculation formula of the zero field specific heat capacity contributed by the CEF is
\begin{equation}
	C_m^{CEF} = \frac{{{N_A}{\partial ^2}\ln \left[ {\sum\limits_{i = 1}^8 {\exp \left( { - \frac{{{E_i}}}{{{k_B}T}}} \right)} } \right]}}{{\partial {{\left( {\frac{1}{{{k_B}T}}} \right)}^2}}}
\end{equation}

For \ce{NaYbSe2}, the CEF $B_{m}^{n}$ parameters are shown in TABLE. III.
\begin{table}
	\caption{CEF $B_{m}^{n}$ parameters for \ce{NaYbSe2}. The units are in meV}
	\begin{ruledtabular}
		\begin{tabular}{cccccc}
			$B_{2}^{0}$ & $B_{4}^{0}$ & $B_{4}^{3}$ & $B_{6}^{0}$ & $B_{6}^{3}$ & $B_{6}^{6}$ \\
			\hline
			-0.1579 & 0.0110 & -0.0999 & 0 & -0.0025 & 0.0133
		\end{tabular}
	\end{ruledtabular}
\end{table}

\section{mean field approximation of anisotropic spin Hamiltonian}
An anisotropic Hamiltonian that has an anisotropic spin interaction and satisfies the symmetry of the R-3m space group (as shown in Fig. 1S(a)) has the following form\cite{Li2015a}:
\begin{equation}
	\hat{H}_{spin-spin} = \sum_{\langle ij\rangle}[J_{zz}S_{i}^{z}S_{j}^{z}+J_{\pm}(S_{i}^{+}S_{j}^{-}+S_{i}^{-}S_{j}^{+}) + J_{\pm\pm}(\gamma_{ij}S_{i}^{+}S_{j}^{+}+\gamma_{ij}^{*}S_{i}^{-}S_{j}^{-})-\frac{iJ_{z\pm}}{2}(\gamma_{ij}S_{i}^{+}S_{j}^{z}-\gamma_{ij}^{*}S_{i}^{-}S_{j}^{z}+\langle i \longleftrightarrow j \rangle)]
\end{equation}
where $J_{zz}, J_{\pm}, J_{\pm\pm}$ and $J_{z\pm}$ are anisotropic spin exchange parameters, the phase factor
$\gamma_{ij}=1, e^{i2\pi/3}, e^{-i2\pi/3}$ for nearnest neighbor (NN) interaction along the $\vec{a}_{1}, \vec{a}_{2}$ and $\vec{a}_{3}$ direction. When a magnetic field is applied, the zeeman term is expressed as follows:
\begin{equation}
	\hat{H}_{zeeman} = -\mu_{0} \mu_{B} \sum_{i}\left[g_{a b}\left(h_{x} S_{i}^{x}+h_{y} S_{i}^{y}\right)+g_{c} h_{c} S_{i}^{z}\right]
\end{equation}

Using the following reference formula, the Hamiltonian can be transformed into the following form for the mean-field approximation.
\begin{equation}
	\begin{array}{l}
	S_{i}^{+} S_{j}^{-}+S_{i}^{-} S_{j}^{+}=2\left(S_{i}^{x} S_{j}^{x}+S_{i}^{y} S_{j}^{y}\right) \\
	S_{i}^{-} S_{j}^{-}=S_{i}^{x} S_{j}^{x}-i S_{i}^{x} S_{j}^{y}-i S_{i}^{y} S_{j}^{x}-S_{i}^{y} S_{j}^{y} \\
	S_{i}^{+} S_{j}^{z}=S_{i}^{x} S_{j}^{z}+i S_{i}^{y} S_{j}^{z} \\
	S_{i}^{-} S_{j}^{z}=S_{i}^{x} S_{j}^{z}-i S_{i}^{y} S_{j}^{z}
	\end{array}
\end{equation}

The transformed Hamiltonian can be written as a matrix expression in different directions.
\begin{equation}
	\hat{H}_{\vec{a}_{1}}=\left(\begin{array}{ccc}
	S_{i}^{x} & S_{i}^{y} & S_{i}^{z}
	\end{array}\right)\left(\begin{array}{ccc}
	2\left(J_{\pm}+J_{\pm \pm}\right) & 0 & 0 \\
	0 & 2\left(J_{\pm}-J_{\pm \pm}\right) & 2 J_{z \pm} \\
	0 & 2 J_{z \pm} & J_{z z}
	\end{array}\right)\left(\begin{array}{c}
	S_{j}^{x} \\
	S_{j}^{y} \\
	S_{j}^{z}
	\end{array}\right)
\end{equation}
\begin{equation}
	\hat{H}_{\vec{a}_{2}}=\left(\begin{array}{ccc}
	S_{i}^{x} & S_{i}^{y} & S_{i}^{z}
	\end{array}\right)\left(\begin{array}{ccc}
	2 J_{\pm}-J_{\pm \pm} & -\sqrt{3} J_{\pm \pm} & -\sqrt{3} J_{z \pm} / 2 \\
	-\sqrt{3} J_{\pm \pm} & 2 J_{\pm}+J_{\pm \pm} & -J_{z \pm 1} / 2 \\
	-\sqrt{3} J_{z \pm} / 2 & -J_{z \pm} / 2 & J_{z=}
	\end{array}\right)\left(\begin{array}{c}
	S_{j}^{x} \\
	S_{j}^{y} \\
	S_{j}^{z}
	\end{array}\right)
\end{equation}
\begin{equation}
	\hat{H}_{\vec{a}_{3}}=\left(\begin{array}{ccc}
	S_{i}^{x} & S_{i}^{y} & S_{i}^{z}
	\end{array}\right)\left(\begin{array}{ccc}
	2 J_{\pm}-J_{\pm \pm} & \sqrt{3} J_{\pm \pm} & \sqrt{3} J_{z \pm} / 2 \\
	\sqrt{3} J_{\pm \pm} & 2 J_{\pm}+J_{\pm \pm} & -J_{z \pm} / 2 \\
	\sqrt{3} J_{z \pm} / 2 & -J_{z \pm} / 2 & J_{z=}
	\end{array}\right)\left(\begin{array}{c}
	S_{j}^{x} \\
	S_{j}^{y} \\
	S_{j}^{z}
	\end{array}\right)
\end{equation}
Do the mean-field approximation for the three directions respectively to get the following expression.
\begin{equation}
\begin{array}{l}
\hat{H}_{\vec{a}_{1}-MF}=\frac{2 J_{z z} m^{z}}{\mu_{0} \mu_{B} g_{c}} \sum_{i} S_{i}^{z}+\frac{4\left(J_{\pm}+J_{\pm \pm}\right) m^{x}}{\mu_{0} \mu_{B} g_{a b}} \sum_{i} S_{i}^{x}+\frac{4\left(J_{\pm}-J_{\pm \pm}\right) m^{y}}{\mu_{0} \mu_{B} g_{a b}} \sum_{i} S_{i}^{y}+\frac{4 J_{z \pm} m^{z}}{\mu_{0} \mu_{B} g_{c}} \sum_{i} S_{i}^{y}+\frac{4 J_{z \pm} m^{y}}{\mu_{0} \mu_{B} g_{a b}} \sum_{i} S_{i}^{z} \\
-\frac{J_{z z} N\left(m^{z}\right)^{2}}{2\left(\mu_{0} \mu_{B} g_{c}\right)^{2}}-\frac{\left(J_{\pm}+J_{\pm \pm}\right) N\left(m^{x}\right)^{2}}{\left(\mu_{0} \mu_{B} g_{a b}\right)^{2}}-\frac{\left(J_{\pm}-J_{\pm \pm}\right) N\left(m^{y}\right)^{2}}{\left(\mu_{0} \mu_{B} g_{a b}\right)^{2}}-\frac{2 J_{z \pm} N m^{y} m^{z}}{\left(\mu_{0} \mu_{B} g_{a b}\right)\left(\mu_{0} \mu_{B} g_{c}\right)}
\end{array}
\end{equation}
\begin{equation}
\begin{array}{l}
\hat{H}_{\vec{a}_{2}-MF}=\frac{2 J_{z z} m^{z}}{\mu_{0} \mu_{B} g_{c}} \sum_{i} S_{i}^{z}+\frac{2\left(2 J_{\pm}-J_{\pm \pm}\right) m^{x}}{\mu_{0} \mu_{B} g_{a b}} \sum_{i} S_{i}^{x}+\frac{2\left(2 J_{\pm}+J_{\pm \pm}\right) m^{y}}{\mu_{0} \mu_{B} g_{a b}} \sum_{i} S_{i}^{y}-\frac{2 \sqrt{3} J_{\pm \pm} m^{y}}{\mu_{0} \mu_{B} g_{a b}} \sum_{i} S_{i}^{x} \\
-\frac{2 \sqrt{3} J_{\pm \pm} m^{x}}{\mu_{0} \mu_{B} g_{a b}} \sum_{i} S_{i}^{y}-\frac{J_{z \pm} \sqrt{3} m^{x}}{\mu_{0} \mu_{B} g_{a b}} \sum_{i} S_{i}^{z}-\frac{J_{z \pm} \sqrt{3} m^{z}}{\mu_{0} \mu_{B} g_{c}} \sum_{i} S_{i}^{x}-\frac{J_{z t} m^{y}}{\mu_{0} \mu_{B} g_{a b}} \sum_{i} S_{i}^{z}-\frac{J_{z \pm} m^{z}}{\mu_{0} \mu_{B} g_{c}} \sum_{i} S_{i}^{y} \\
-\frac{J_{z z} N\left(m^{z}\right)^{2}}{2\left(\mu_{0} \mu_{B} g_{c}\right)^{2}}-\frac{\left(2 J_{\pm}-J_{\pm \pm}\right) N\left(m^{x}\right)^{2}}{2\left(\mu_{0} \mu_{B} g_{a b}\right)^{2}}-\frac{\left(2 J_{\pm}+J_{\pm \pm}\right) N\left(m^{y}\right)^{2}}{2\left(\mu_{0} \mu_{B} g_{a b}\right)^{2}}+\frac{2 \sqrt{3} J_{\pm \pm} N m^{y} m^{x}}{2\left(\mu_{0} \mu_{B} g_{a b}\right)^{2}}+\frac{J_{z \pm} \sqrt{3} N m^{x} m^{z}}{2\left(\mu_{0} \mu_{B} g_{a b}\right)\left(\mu_{0} \mu_{B} g_{c}\right)}+\frac{J_{z \pm} N m^{y} m^{z}}{2\left(\mu_{0} \mu_{B} g_{a b}\right)\left(\mu_{0} \mu_{B} g_{c}\right)}
\end{array}
\end{equation}
\begin{equation}
\begin{array}{l}
\hat{H}_{\vec{a}_{3}-MF}=\frac{2 J_{z z} m^{z}}{\mu_{0} \mu_{B} g_{c}} \sum_{i} S_{i}^{z}+\frac{2\left(2 J_{\pm}-J_{\pm \pm}\right) m^{x}}{\mu_{0} \mu_{B} g_{a b}} \sum_{i} S_{i}^{x}+\frac{2\left(2 J_{\pm}+J_{\pm \pm}\right) m^{y}}{\mu_{0} \mu_{B} g_{a b}} \sum_{i} S_{i}^{y}+\frac{2 \sqrt{3} J_{\pm t} m^{y}}{\mu_{0} \mu_{B} g_{a b}} \sum_{i} S_{i}^{x} \\
+\frac{2 \sqrt{3} J_{\pm \pm} m^{x}}{\mu_{0} \mu_{B} g_{a b}} \sum_{i} S_{i}^{y}+\frac{J_{z \pm} \sqrt{3} m^{x}}{\mu_{0} \mu_{B} g_{a b}} \sum_{i} S_{i}^{z}+\frac{J_{z \pm} \sqrt{3} m^{z}}{\mu_{0} \mu_{B} g_{c}} \sum_{i} S_{i}^{x}-\frac{J_{z_{i}} m^{y}}{\mu_{0} \mu_{B} g_{a b}} \sum_{i} S_{i}^{z}-\frac{J_{z \pm} m^{z}}{\mu_{0} \mu_{B} g_{c}} \sum_{i} S_{i}^{y} \\
-\frac{J_{z z} N\left(m^{z}\right)^{2}}{2\left(\mu_{0} \mu_{B} g_{c}\right)^{2}}-\frac{\left(2 J_{\pm}-J_{\pm \pm}\right) N\left(m^{x}\right)^{2}}{2\left(\mu_{0} \mu_{B} g_{a b}\right)^{2}}-\frac{\left(2 J_{\pm}+J_{\pm \pm}\right) N\left(m^{y}\right)^{2}}{2\left(\mu_{0} \mu_{B} g_{a b}\right)^{2}}-\frac{2 \sqrt{3} J_{\pm \pm} N m^{y} m^{x}}{2\left(\mu_{0} \mu_{B} g_{a b}\right)^{2}}-\frac{J_{z \pm} \sqrt{3} N m^{x} m^{z}}{2\left(\mu_{0} \mu_{B} g_{a b}\right)\left(\mu_{0} \mu_{B} g_{c}\right)}+\frac{J_{z \pm} N m^{y} m^{z}}{2\left(\mu_{0} \mu_{B} g_{a b}\right)\left(\mu_{0} \mu_{B} g_{c}\right)}
\end{array}
\end{equation}
Where $m_{x}$, $m_{y}$, $m_{z}$ are the magnetic moments corresponding to the three spin orientations of $x$, $y$, and $z$ respectively.

We add a magnetic field to the ab-plane and select the polarization direction as $S_{x}$, then $m^{x}=M^{x}$, $m^{y}=0$, $m^{z}=0$. The expression of the mean field Hamiltonian of the ab-plane is as follows:
\begin{equation}
\begin{array}{l}
\hat{H}_{MF-ab}=\hat{H}_{\vec{a}_{1}}+\hat{H}_{\vec{a}_{2}}+\hat{H}_{\vec{a}_{3}}-\mu_{0} \mu_{B} g_{ab} h_{ab} \sum_{i} S_{i}^{x} \\
=\left(\frac{12 J_{\pm} M^{x}}{\mu_{0} \mu_{B} g_{a b}}-\mu_{0} \mu_{B} h_{a b}\right) \sum_{i} S_{i}^{x}+\frac{\left(2 J_{\pm \pm}-J_{\pm}\right) N\left(M^{x}\right)^{2}}{\left(\mu_{0} \mu_{B} g_{a b}\right)^{2}}
\end{array}
\end{equation}

We add a magnetic field to the c-axis and select the polarization direction as $S_{z}$, then $m^{x}=0$, $m^{y}=0$, $m^{z}=M^{z}$. The expression of the mean field Hamiltonian of the ab-plane is as follows:
\begin{equation}
\begin{array}{l}
\hat{H}_{MF-c}=\hat{H}_{\vec{a}_{1}}+\hat{H}_{\vec{a}_{2}}+\hat{H}_{\vec{a}_{3}}-\mu_{0} \mu_{B} g_{c} h_{c} \sum_{i} S_{i}^{z} \\
=\left(\frac{6 J_{z z} m^{z}}{\mu_{0} \mu_{B} g_{c}}-\mu_{0} \mu_{B} g_{c} h_{c}\right) \sum_{i} S_{i}^{z}+\frac{2 J_{z \pm} m^{z}}{\mu_{0} \mu_{B} g_{c}} \sum_{i} S_{i}^{y}-\frac{3 J_{z z} N\left(m^{z}\right)^{2}}{2\left(\mu_{0} \mu_{B} g_{c}\right)^{2}}
\end{array}
\end{equation}

\begin{figure}[t]
	\includegraphics[scale=0.7]{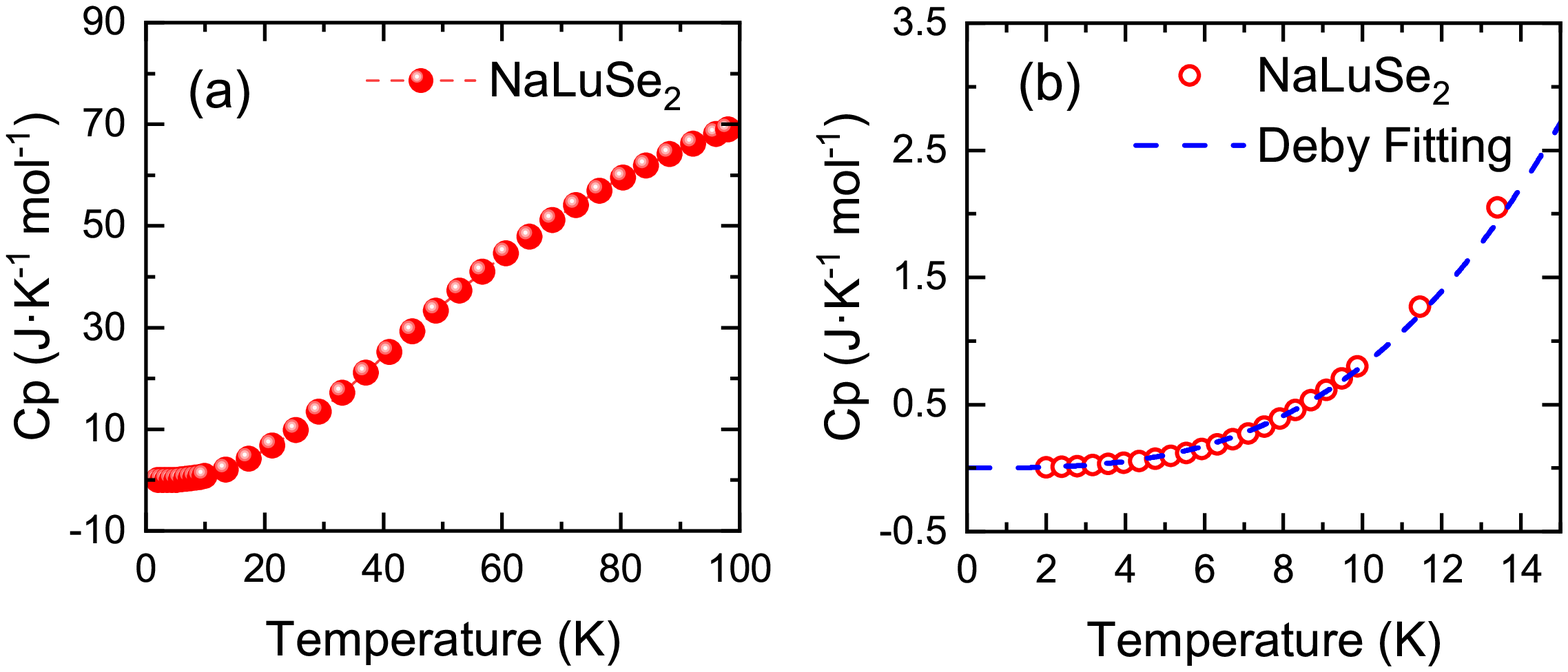}
	\caption{\label{fig:epsart}(a) Zero field specific heat capacity data of NaLuSe2. (b) Low temperature Debye fitting of \ce{NaLuSe2}}
\end{figure}

Combining the thermodynamic calculation formula and expression (24) and expression (25), we can calculate the magnetic susceptibility and magnetization in the ab-plane and along the c-axis directions, respectively.

\section{specific heat capacity of \ce{NaLuSe2}}

As shown in Fig. S2(b), the low-temperature specific heat capacity of \ce{NaLuSe2} can be well fitted by the Debye formula. Therefore, \ce{NaLuSe2} can be used as a good non-magnetic control sample to simulate the contribution of phonon heat capacity of \ce{NaYbSe2}.

\section{MagLab software package}
In order to analyze the crystal field of rare-earth ion compounds more conveniently, we have developed a python-based software package MagLab.
As shown in the Fig. S3, the MagLab software package consists of the following 7 modules. 

Data Module, LoadData Module and PhyUnitesAndCons Module respectively correspond to the green squares of Fig. S3 from left to right. These three modules are responsible for the storage and preprocessing of the original experimental data, the loading of the original experimental data and the conversion of related physical units. The CrystalField Module (red sqaure in Fig. S3) is the core of CEF data analysis, mainly used to construct CEF Hamiltonian with different crystallographic symmetry structures. The Thermodynamics module (left blue square in Fig. S3) is responsible for data calculation of CEF magnetic susceptibility, CEF magnetization, and heat capacity contribution of CEF. The Optimization Module contains two optimization algorithms, gradient descent algorithm and simulated annealing algorithm, which can fit the CEF parameters according to the loaded experimental data. At the same time, the optimization module is complementary to the thermodynamics module, and the two can cooperate with each other to achieve combination of experimental data and theoretical analysis method. Log Module is responsible for saving calculation results.
\begin{figure}[t]
	\includegraphics[scale=1]{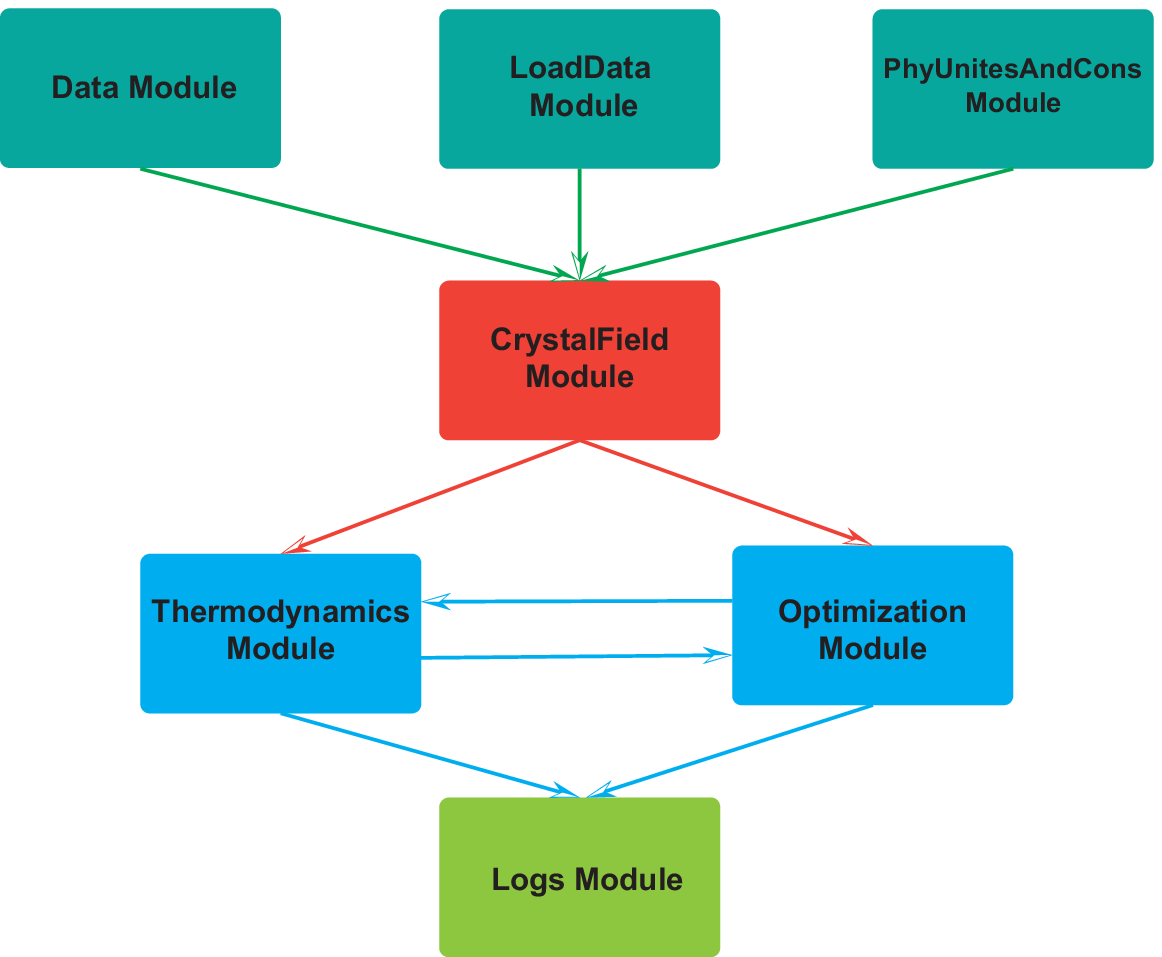}
	\caption{\label{fig:epsart}Module organization structure of MagLab software package}
\end{figure}

Based on the MagLab package, we have calculated the magnetic susceptibility and heat capacity of the \ce{NaYbSe2} CEF, and the calculated results are exactly the same as those given by the Mantid\cite{Arnold2014} software. We will continue to improve the functions of MagLab and expand other magnetic-related calculation methods.


\emph{Acknowledgments.}---This work was supported by the National Key Research and Development Program of China (2017YFA0302904 \& 2016YFA0300500) and the NSF of China (11774419 \& U1932215). A portion of this work was performed on the Steady High Magnetic Field Facilities, High Magnetic Field Laboratory, CAS. The CEF calculations of magnetic susceptibility is based on our python program package MagLab, which is designed for the fitting, calculations, and analysis of the CEF magnetism, and will be continuously updated. The MF analysis and fitting of magnetic susceptibility, magnetization and ESR data at low temperatures are based on MathWorks$^\circledR$ MATLAB$^\circledR$ software (Academic License for Renmin University of China).


%